\documentclass[aps,pra,floatfix,showpacs,tightenlines,twocolumn,superscriptaddress]{revtex4}

\usepackage{graphicx}
\usepackage{epsfig}

\begin{document}

\title{Quantum Computation by Communication}

\author{T. P. Spiller}\email{timothy.spiller@hp.com}
\affiliation{Hewlett-Packard Laboratories, Filton Road, Stoke
Gifford, Bristol BS34 8QZ, United Kingdom}

\author{Kae Nemoto}\email{nemoto@nii.ac.jp}
\affiliation{National Institute of Informatics, 2-1-2 Hitotsubashi,
Chiyoda-ku, Tokyo 101-8430, Japan}

\author{Samuel L. Braunstein}
\affiliation{Computer Science, University of York, York YO10 5DD,
United Kingdom}

\author{W. J. Munro}
\affiliation{National Institute of Informatics, 2-1-2 Hitotsubashi,
Chiyoda-ku, Tokyo 101-8430, Japan} \affiliation{Hewlett-Packard
Laboratories, Filton Road, Stoke Gifford, Bristol BS34 8QZ, United
Kingdom}

\author{P. van Loock}
\affiliation{National Institute of Informatics, 2-1-2 Hitotsubashi,
Chiyoda-ku, Tokyo 101-8430, Japan}

\author{G. J. Milburn}
\affiliation{Centre for Quantum Computer Technology,Department of
Physics, University of Queensland, Australia}

\date{\today}

\begin{abstract}
We present a new approach to scalable quantum computing---a ``qubus
computer''---which realises qubit measurement and quantum gates
through interacting qubits with a
quantum communication bus mode. The qubits could be ``static''
matter qubits or ``flying'' optical qubits, but the scheme we focus
on here is particularly suited to matter qubits. There is no
requirement for direct interaction between the qubits. Universal
two-qubit quantum gates may be effected by schemes which involve
measurement of the bus mode, or by schemes where the bus disentangles
automatically and no measurement is needed.
In effect, the approach integrates together
qubit degrees of freedom for computation with quantum continuous
variables for communication and interaction.
\end{abstract}


\maketitle

\section{Introduction}

Quantum computing has reached a very interesting stage in its
development. Over the last decade there have been numerous proposals
for qubit realisations \cite{fort00}. Some of the more mature
proposals, such as trapped ions \cite{cirzol95}, nuclear spins (in
molecules in liquid state) \cite{nmr97} and photonic qubits
\cite{klm01} have now been demonstrated to work in the laboratory at
the few-qubit level. Now in terms of their long term prospects for
scalability, there is deemed to be considerable promise in ``solid
state'' qubits, based (directly or indirectly) on fabrication and
technologies developed for conventional IT. However, at present such
approaches lag behind the more mature ones--- they are either still
on the drawing board, or at the one- or two-qubit demonstration
level. The promise of scalability has yet to achieved
for these approaches, and over the next few years it will be
interesting to see which systems can meet this challenge and which
founder.

Clearly decoherence and measurement are both important and
challenging issues for solid state qubits. With the current
emergence of demonstration qubit experiments, there is optimism
about these problems being solved to a level that would permit
useful small-scale quantum processing. However, even if these
problems can be solved, there is still a need for two-qubit quantum
gates to be implemented in a manner that enables the addition of
more qubits to a system, so there is scalability. This is the main
issue that we address in this paper. If these gates are implemented through
a direct qubit-qubit interaction (i.e. a direct qubit-qubit
coupling term in the basic system Hamiltonian),
potential problems with two-qubit
gates are: (i) the addition of an extra qubit to a system may
disrupt the settings and calibrations that have been put in place
for quantum computing with the original system, and (ii) the qubits
may have to be so close together that individual addressing (both
for single-qubit gates and measurement) cannot be achieved. Direct
qubit interactions may be just fine for demonstrating entanglement
between two solid state qubits, but they may not be as good when it
comes to building a universal and scalable quantum processor.
For example, with just nearest neighbour interactions
there is a large SWAP operation overhead to interact chosen qubits,
that could be removed through use of a bus to mediate interactions
between non-nearest neighbours. One well known technique is to use
single photons to mediate this interaction. There have been a number
of very elegant proposals focusing on this, but they place
highly stringent requirements, for example, on the generation of the single
photons or their detection \cite{cir97,cir99,mancini04,duan04,lim05,duan05,barrett05,refwithin}.

The approach that we present here contains no direct
qubit-qubit interactions and does not require the use of single
photons. Such interactions are achieved indirectly
through the interaction of qubits with a common quantum field
mode---a continuous quantum variable (CV) \cite{bra05,bra92}---which
can be thought of
as a communication bus \cite{lloyd00} . Our ``qubus computer'' approach brings
together the best of both worlds. Static solid state qubits are used
where they work best, for processing. Continuous variables are used
where they work best, for communication and mediating interactions;
they also have the potential to enable interfacing with existing,
conventional information technology. We will assume that
individual qubits can be prepared, subjected to single-qubit
operations and measured. However, as we discuss in order to
introduce our approach, interaction of a qubit with a CV bus mode,
followed by measurement of the bus mode, can also be used in order to
effect quantum non-demolition (QND) measurement of the qubit. This could be the
preferred measurement scheme, unless something even better is
achievable by other means. Our approach is based on qubits
interacting with the bus mode through distinct dipole couplings,
such as the electric dipole of a charge qubit coupling to the
electric field quadrature, or the dipole of a spin
or magnetic moment coupling to the magnetic field quadrature. The
approach should be widely applicable in the solid state qubit
context and so we present it in a generic fashion without being
tied to any specific implementations.

As will be seen, our whole approach is based on the idea of a
sequence of interactions, or gates, between qubits and the bus mode,
followed by measurement of the bus in some scenarios, and not in others.
The concept is therefore that qubits can be brought into interaction
with the bus mode to effect the desired gate sequence, or that (certainly
in the scenarios which involve bus measurement) a bus mode
pulse can be employed
to interact with successive qubits to effect the gate sequence.
Our approach is thus to be contrasted with an ``always on'' interaction
between qubits and a bus mode. In the latter case,
such an interaction can in effect
mimic a direct qubit-qubit coupling. For example, two qubits
simultaneously coupled to a bus through the Jaynes-Cummings interaction
behave as if they have a direct exchange interaction in the dispersive
limit \cite{zhe00,ger05}. Instead, in our approach the
qubit-bus interactions are
sequential.

Following the generic formulation of our approach, we give an illustration
applicable to superconducting charge qubits \cite{shn97,nak99}. The main
results we present are methods for performing universal two-qubit gates
mediated through a CV bus. Now in the superconducting scenario approaches
have been proposed for using a common oscillator mode to effect qubit
interactions \cite{mak99,bla04}, in effect by mimicking direct qubit-qubit
interaction.  It is also possible to consider
an analogy with ion traps \cite{cirzol95} or cavity
QED \cite{zheng04} for such solid state systems \cite{liu05}.
Here we present a variety  of schemes for two-qubit gates,
which utilize a sequence of qubit interactions with the common bus mode
in different ways, and which should be applicable to a wide range of
matter qubit systems.

One approach we give requires no post-interaction work
on the bus mode---it disentangles automatically from the qubits when
the gate is done. Such schemes are analogous to ion trap gates which are insensitive
to the vibrational state of the ions \cite{mil99,mol99,sor00,mil00}.
This form of gate probably has the most widespread promise
and long-term potential. However, it is also possible to effect
gates that require a post-interaction measurement of the bus mode, based
on recent ideas from non-linear quantum optics \cite{nem04,mun05b,barrett05a,mun05a,barrett05b}.
These schemes may be preferable for some systems (certainly so if all the qubits couple
to the same quadrature of the bus), and may also be the simplest approach for
initial experimental investigations. They are also the natural extension of
the QND measurement approach applied to two qubits, so we include
discussion of various schemes of this form.
In addition, the bus-measurement-free approach may be
applicable in the field of non-linear quantum optics (or, more
generally, where the interaction Hamiltonian has the characteristic
cross-Kerr form), so we also include a discussion of this here.

We describe our qubits using the conventional Pauli operators, with
the computational basis being given by the eigenstates of
$\sigma_z$, with $|0\rangle \equiv |\uparrow_z\rangle$ and $|1\rangle
\equiv |\downarrow_z\rangle$. The communication bus mode is
described as a quantum field mode with creation(annihilation)
operators $a^{\dagger}(a)$. For many solid-state qubits this could
be an electromagnetic microwave field mode, although for other
systems it may be an optical field. The centrepiece of our approach
is an interaction Hamiltonian of the form
\begin{equation}
\label{Hint} H_{int} = \hbar \chi \sigma_z X(\theta)
\end{equation}
where $\sigma_z$ is the qubit operator and the field quadrature
operator is $X(\theta)=(a^{\dagger} e^{i \theta} + a e^{-i
\theta})$.
Such an interaction Hamiltonian arises from the interaction between
a charge qubit
or Cooper pair box \cite{shn97,nak99} and the electromagnetic field, with
the $z$ eigenstates representing the relevant two excess charge
states of the box or island. Further examples include the interaction
of a Cooper-pair box with a micromechanical resonator or cantilever
\cite{arm02}, and other quantum electromechanical systems such as
a Fullerene quantum dot which can both carry excess charge and vibrate
mechanically \cite{wah04}. All of these systems can exhibit a very large
electric dipole moment (compared to traditional atomic systems) and
couple strongly to the relevant oscillator or field mode.
The action of the Hamiltonian (\ref{Hint}) for a time $t$ effects a
displacement operation on the field of $D(\sigma_z \beta)$ \cite{inter-pic},
conditioned on the state of the qubit, where $\beta = \chi t
e^{i(\theta - \frac{\pi}{2})}$ and $D$ is the usual displacement
operator $D(\beta) = \exp(\beta a^{\dagger} - \beta^* a)$.

\section{Qubit measurement through controlled displacement}

As an introduction to the use of a coherent bus mode we consider
its application for measurement of a qubit. For the case of $\theta
= \pi/2$ (coupling to the momentum quadrature in Eq.~(\ref{Hint})),
the displacements are in the $X(0)$ direction and, after
interaction, an initial qubit-bus state of $|\Psi_i\rangle = (c_0
|0\rangle + c_1 |1\rangle)|\alpha\rangle$ is transformed to the
entangled state
\begin{equation}
|\Psi_f\rangle = c_0 |0\rangle |\alpha + \beta\rangle + c_1
|1\rangle |\alpha - \beta\rangle \; .
\end{equation}

\begin{figure}[!htb]
\begin{center}\includegraphics[scale=0.8]{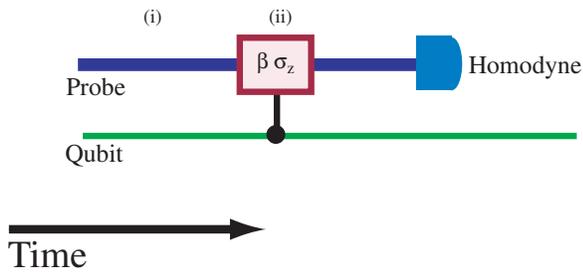}\end{center}
\caption{Circuit diagram for the QND measurement scheme based on
conditional displacement of the bus mode by the qubit followed by a homodyne
measurement on the bus mode.}
\label{measure}
\end{figure}

\begin{figure}[!htb]
\begin{center}\includegraphics[scale=0.45]{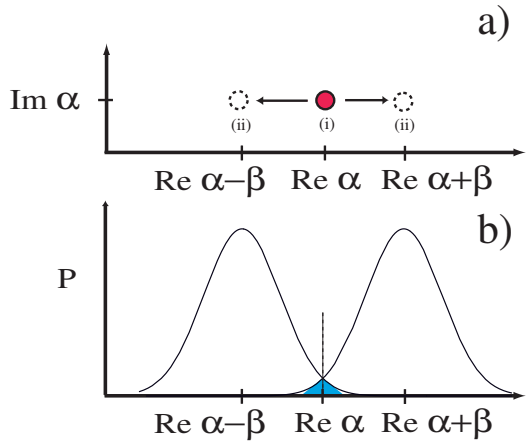}\end{center}
\caption{Schematic diagram showing the effect on the CV bus coherent state of
interaction with the qubit (a). The coherent state is displaced conditional on the
qubit state. The labels (i) and (ii) correspond to the evolution stages, as shown in the
circuit diagram in the previous figure. (b) A schematic diagram of the
$X(0)$ quadrature probability distribution
for the CV bus mode state after interaction with the qubit. Clearly the qubit
measurement is not perfect due to the overlap of peaks corresponding to
different computational basis states.}
\label{error}
\end{figure}

The circuit diagram for this is shown in Fig.~\ref{measure} for $\beta$ real.
The effect of the interaction on the CV bus
in phase space is illustrated in Fig.~\ref{error}.
 Measurement of the bus mode can thus effect a non-demolition
measurement of the qubit in its computational basis. The field
measurement could be made by a homodyne measurement of the $X(0)$
quadrature, or an intensity measurement. This approach is the
displacement-based analogue of photon non-demolition measurement
based on an optical cross-Kerr non-linearity
\cite{imo85,mil84,mun05}. Homodyne measurement of the bus mode,
for example if this is a microwave field mode coupled to matter
qubits, may be effected through a single electron transistor
operated as a mixer
\cite{saro05}, or some other suitable non-linear device,
such as a superconducting Josephson ring system \cite{maria05}.


Now clearly the qubit measurement is not perfect, as
the final states of the CV bus mode corresponding to the
different computational basis state amplitudes for the qubit
are not exactly orthogonal, as illustrated in Fig.~\ref{error}(b).
However, for the example of taking the midpoint between the
probability peaks as the discrimination point and using the $X(0)$
quadrature measurement of the bus, the error probability (the sum of
the areas that sit the wrong side of the discrimination point) is
$E = \frac{1}{2}$erfc$(2^{-1/2} |\beta|)$. This can be made very small
for a suitable choice of $\beta$ \cite{mun05}, for example even for
$|\beta| \sim 3$ the error is $E \sim 0.001$. Now this error formula and the
example numbers are given on the basis of very accurate homodyne
measurement of the CV mode quadrature. However, even
somewhat imperfect homodyne measurement would still provide very good qubit
measurement. The usual way to describe such an imperfect measurement is via a
gaussian convolution of the ideal homodyne projector\cite{Wiseman93,Tyc04}, that is we use the
projector
$(2\pi\Delta)^{-1/2}\int_{-\infty}^\infty dy \exp[-(X-y)^2/(2\Delta)]|y\rangle \langle y|$
instead of $|X\rangle \langle X|$. The effect of this is to broaden the two distributions
from a width of unity to a width of $1+\Delta$, which in turn means the
overlap error function $E$ changes by a rescaling of $\beta$ to
$\frac{\beta}{\sqrt{1+\Delta}}$. This can clearly still be kept small
for a suitable choice of $\beta$.

This CV bus approach to solid state qubit measurement clearly has much promise.
For example, it has already been realised \cite{wal05} for a superconducting charge
qubit coupled to a microwave mode in the dispersive limit, when the
qubit-cavity coupling effectively takes the form of a cross-Kerr
non-linearity \cite{bla04,wal04} rather than that which generates
controlled displacements.

\section{Two-qubit interaction through controlled bus
displacement and measurement}

Measurement of a coherent bus mode, following its interaction with
two qubits, can be used to effect an entangling operation between
the qubits. As an example, we again consider displacements in the
$X(0)$ direction of the field. After the interactions, an initial
two-qubit-bus product state of
\begin{equation}
\label{Psiinitial}
 |\Psi_i\rangle = \frac{1}{2}(|00\rangle +
|01\rangle + |10\rangle +  |11\rangle)|\alpha\rangle
\end{equation}
 is transformed to
\begin{equation}
\label{twoqubitbus}
|\Psi_f\rangle = \frac{1}{2} \left( |00\rangle |\alpha + 2
\beta\rangle +  (|01\rangle + |10\rangle)|\alpha\rangle  +
|11\rangle |\alpha - 2 \beta \rangle \right),
\end{equation}
assuming equal strength coupling of both qubits to the bus mode. Now if
this procedure can be performed with the vacuum state ($\alpha =
0$), then all is well and good. If not, an {\it unconditional}
displacement operation $D(-\alpha)$ is applied to the bus mode prior
to measurement. With this resolved, an appropiate measurement of the
bus mode projects the two-qubit system into a maximally entangled
state.

\begin{figure}
\begin{center}\includegraphics[scale=0.75]{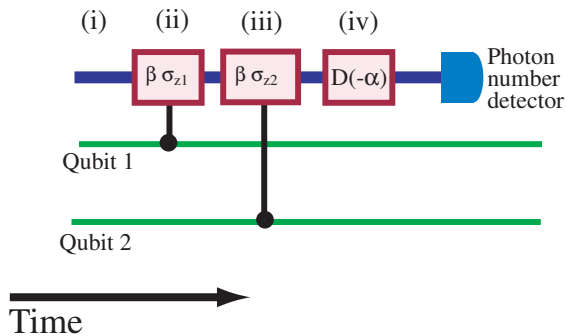}\end{center}
\caption{Circuit diagram for a two-qubit parity gate based on
controlled displacements between the qubits and the ``probe'' bus,
followed by bus measurement.}
\label{dispandmeascirc}
\end{figure}

\begin{figure}
\begin{center}\includegraphics[scale=0.45]{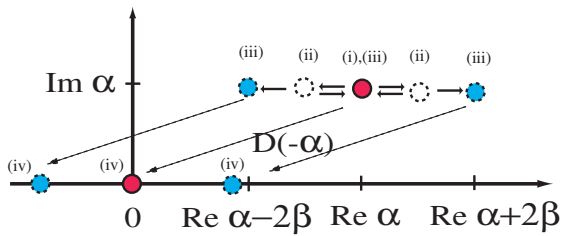}\end{center}
\caption{Schematic diagram showing the phase space evolution of the bus amplitudes
corresponding to the various qubit computational basis states. The
labels (i) to (iv) correspond to the evolution stages, as shown in the
circuit diagram in Fig \ref{dispandmeascirc}.}
\label{dispandmeas}
\end{figure}

The corresponding circuit diagram for this operation
is shown in Fig.~\ref{dispandmeascirc}. The evolution of the CV bus mode
amplitudes is illustrated in Fig.~\ref{dispandmeas}.
Now, for an ideal projection onto $|n\rangle$\cite{projection}, the two-qubit state
is conditioned to
\begin{eqnarray}
|\psi_f\rangle = 2^{-1/2}(|01\rangle + |10\rangle)
\label{vacresult}
\end{eqnarray}
for $n=0$ and to
\begin{eqnarray}
|\psi_f\rangle = 2^{-1/2}(|00\rangle + (-1)^n |11\rangle)
\label{nresult}
\end{eqnarray}
for $n>0$. These projections happen with equal probability of $\frac{1}{2}$
(with the most likely value of $n$ in the latter case being $n \sim
4|\beta|^2$), although there is an error probability of
$e^{-4|\beta|^2}$ (and a corresponding admixture to the state) in
the former case, due to the overlap of the coherent state with the
vacuum. This error can be made small for a suitable choice of
$\beta$. The phase factor in the $n>0$ case is heralded by the
measurement outcome $n$, and so can be allowed for or corrected.

An alternative approach, which may be preferable in initial experiments, is
to use an intensity measurement (or even a bucket detector),
instead of a photon number resolving detector. In such a case the odd parity state is
still projected to $|01\rangle + |10\rangle$, as in Eq.~(\ref{vacresult}).
However the even parity state of Eq.~(\ref{nresult})
$|00\rangle + (-1)^n |11\rangle$ becomes mixed, due to the uncertainty of whether $n$,
although $>0$, is
even or odd. The even parity state is thus represented by the density matrix
$\rho= A \left[|00\rangle + |11\rangle\right]\left[\langle 00|+\langle 11|\right]
+(1-A) \left[|00\rangle - |11\rangle\right]\left[\langle 00|-\langle 11|\right]$
where $A$ is determined by the distribution of the amplitudes $|00\rangle +|11\rangle$ and
$|00\rangle -|11\rangle$ in the even parity subspace and any information obtained
about $n$ from the measurement \cite{bucketdet}.
Now even in the worst case scenario ($A=1/2$) the protocol can be repeated, applying
Hadamard operations ($|0\rangle \rightarrow |0\rangle + |1\rangle \; , \;
|1\rangle \rightarrow |0\rangle - |1\rangle$) to the qubits and then
interacting with the bus and measuring. Half the time, and heralded, this will give the
odd parity pure state of Eq.~(\ref{vacresult}) (which could be deterministically
transformed to some other entangled state, as desired), and half the time a mixture
will result. Thus after $m$ iterations the level of mixture will be
$\sim (1/2)^m$,  which can be made arbitrarily small by increasing $m$.
Thus with multiples uses of a simple intensity measurement or bucket detector
it is possible to generate a near-deterministic entangling operation between
qubits, without the need for photon number resolution in the detection
device applied to the bus \cite{clusterprep}.

The entangling operation given in Eqs.~(\ref{vacresult}) and (\ref{nresult}),
when operated with photon number resolution on the measurement,
effectively projects the initial two-qubit
state into an odd or even parity entangled state and, since the
outcome is heralded by the measurement result, one could be
transformed to the other, as desired. Whilst such a parity operation is
not a unitary operation, it is possible to utilize this form
of qubit parity operation along with single qubit rotations to construct a
universal gate set \cite{nem05}. This has been shown in
detail in the analogous case for optical qubits
coupled to bus modes through cross-Kerr non-linearities \cite{nem04,mun05a}.
This analysis carries over in a straightforward manner to
similar parity operations, however they are achieved.
So, the displacement-based parity operation presented here provides a
route to universal quantum processing for solid state qubits, all
dipole-coupled to the same quadrature of a bus mode. It is also
worth noting that under certain conditions the coupling of, for
example, a charge qubit to a microwave field behaves like a
cross-Kerr coupling, and so generates controlled rotations rather
than controlled displacements on the field \cite{bla04}. In this
limit the quantum optical approach to gates \cite{nem04,mun05b,barrett05a,mun05a} carries
over directly.

A further point to consider from the perspective of initial experiments
is that it may be much easier to effect a probabilistic (but good fidelity)
entangling operation, rather than the full parity gate. For example, homodyne
measurement of the $X(0)$ quadrature of the CV bus mode applied directly to a
system in the state of Eq.~(\ref{twoqubitbus}) will generate the two-qubit
state of Eq.~(\ref{vacresult}) probabilistically,
but---very importantly---heralded by the quadrature result.
As with the qubit measurement example, provided that $\beta$ is sufficiently
large so the bus state probability distributions corresponding to the
different two-qubit amplitudes in Eq.~(\ref{twoqubitbus}) have very little overlap,
even a somewhat imperfect homodyne measurement of the bus quadrature can
still give very high fidelity two-qubit entanglement. This probabilistic but
heralded entangling operation is a good initial goal for experiments, prior
to the full parity operation, leading further to a universal two-qubit gate.

\section{Two-qubit gate through controlled bus displacements alone}

Now there may be situations, such as initial experimental
tests, or cases where in practice coupling to only one bus
quadrature is possible, where it is desirable to effect a gate
through qubit-bus interactions followed by a bus measurement. However,
it is in fact possible to construct a universal two-qubit gate
purely through a sequence of qubit-bus interactions, {\it without}
the need for any subsequent measurement \cite{mil99,mil00,wan01}. This clearly
simplifies the procedure, but it also has the potential for making
the gate faster, as there is no need for measurement of the bus and
qubit operations conditional on this result to complete the gate.
One scheme to achieve such a
gate requires one qubit (labelled 1) to be coupled to the momentum
quadrature of the field ($\theta = \pi/2$ in Eq.~(\ref{Hint})), thus
generating displacements on the bus in the $X(0)$ direction, and the
other qubit (labelled 2) to be coupled to the position quadrature
($\theta = 0$ in Eq.~(\ref{Hint})), giving displacements in the
orthogonal direction.

\begin{figure}
\begin{center}\includegraphics[scale=0.6]{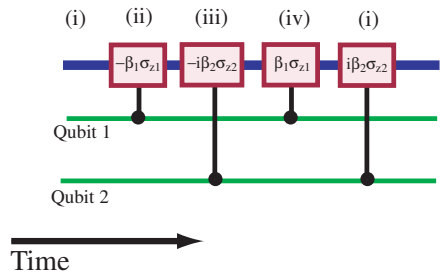}\end{center}
\caption{Circuit diagram of a universal two-qubit gate based on
controlled displacements between the qubits and the ''probe'' bus.}
\label{displacementscirc}
\end{figure}

\begin{figure}
\begin{center}\includegraphics[scale=0.35]{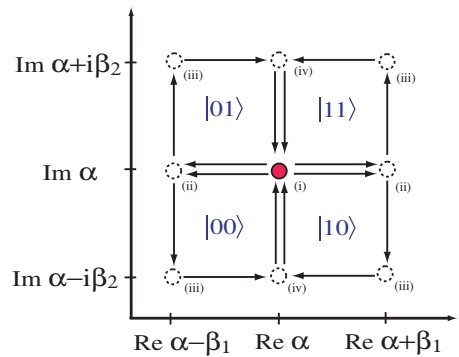}\end{center}
\caption{Schematic diagram showing the phase space evolution of the bus amplitudes
for the four basis states of the two qubits. The
labels (i) to (iv) correspond to the evolution stages, as shown in the
circuit diagram in Fig. \ref{displacementscirc}.}
\label{displacements}
\end{figure}

With this arrangement and using the well-known result
\begin{eqnarray}
D(\beta_1) D(\beta_2)= \exp \left[ \frac{\beta_1 \beta_{2}^\ast -
\beta_{1}^* \beta_2}{2} \right] D(\beta_1 +\beta_2) \; ,
\end{eqnarray}
the gate follows from four conditional displacements. The sequence
of operations is shown in Fig.~\ref{displacementscirc}. This defines
the unitary operator
\begin{eqnarray}
U_{12}(\beta_1,\beta_2) &=& D(i\beta_2 \sigma_{z2}) D(\beta_1
\sigma_{z1}) \nonumber \\
&\;&\;\;\;\;\;\;\;\times D(-i\beta_2 \sigma_{z2}) D(-\beta_1 \sigma_{z1}) \;
\end{eqnarray}
For the case of real $\beta_1$ and $\beta_2$ the effect of this
operator on the bus coherent state, conditional on the state of the
qubits, is illustrated  in Fig.~\ref{displacements}. The action on
the initial state of Eq.~(\ref{Psiinitial}) is
\begin{eqnarray}
|\Psi_f\rangle &=& U_{12}(\beta_1,\beta_2) |\Psi_i\rangle \nonumber \\
 &=& \frac{1}{2} \left( |00\rangle e^{2 i \beta_1 \beta_2}+  |01\rangle e^{-2 i \beta_1 \beta_2} \right. \nonumber \\
&\;&\;\;\;\left. + |10\rangle e^{-2 i \beta_1 \beta_2}  + |11\rangle e^{2 i
\beta_1 \beta_2} \right) |\alpha\rangle \; .
\end{eqnarray}
For any real value of $\beta_1 \beta_2$ the bus mode is disentangled from
the qubits at the end of the operation; nevertheless, for $2 \beta_1
\beta_2 = \pi/4$ we obtain a maximally entangled state of the two
qubits. In this case we achieve a universal two-qubit gate, which is
equivalent to a controlled-phase gate (up to application of local
unitaries $U_i = 2^{-1/2}(1 - i \sigma_{zi})$ to each qubit and a
global phase of $\pi/4$). The bus mode enables the gate to be
performed---it is certainly entangled with the qubits during the
operation---but at the end of the gate it is disentangled and
so has effectively played the role of a catalyst.


There are clearly numerous variations that can be made within this
framework \cite{wan01}. The key features of the gate are:
\begin{enumerate}
\item It is the total
phase space area traced out by a coherent state amplitude that
determines the phase acquired by that amplitude \cite{wan01}.
For a closed anticlockwise path $C$ in phase
space
%\begin{equation}
%\oint_C D(d\alpha) = \exp(2 i \cal{A}_C) \; ,
%\end{equation}
%where $\cal{A}_C$ is the area enclosed by $C$.\begin{equation}
\begin{eqnarray}
\lim_{\Delta \alpha \rightarrow 0}  \Theta \left(\prod_i D(\Delta \alpha_i)\right)= \exp(2 i \cal{A}_C) \; ,
\end{eqnarray}
where $\Theta ()$ reminds us that the operator order is preserved and the
path $\{\Delta \alpha_1,\Delta \alpha_2,\dots   \}$ forms a closed
anticlockwise path $C$ in phase space. $\cal{A}_C$ is the area enclosed
by $C$.
 \item The fact that all the coherent state amplitudes end up
on top of each other at the end of the gate disentangles the bus
from the qubits without
the need for any measurement of the bus mode.
\end{enumerate}
There is a lot of
freedom available within the constraints of achieving these
features. For example, it may well be desirable to work with
$\beta_1 = \beta_2$ (so $\beta_1=(\pi/8)^{1/2}$ achieves the
maximally entangling gate) or thereabouts, in order to minimise the
total displacement for a given area. However, this is not
necessary---different forms of qubit that couple to the bus mode
with different strengths can be used, giving rectangular paths in
phase space. Furthermore, the displacements do not have to be in
orthogonal directions, although clearly a greater total displacement
distance is required to achieve a given gate (such as a maximally
entangling benchmark) if non-orthogonal displacements are employed.
In general the shapes of the paths in phase space don't matter;
what is essential is that different two-qubit amplitudes effect
different closed path areas on the bus, so that the phases
acquired generate an entangling gate. In this sense the gate can be
regarded as a geometric phase gate \cite{wan01}.

Although we have illustrated the gate with coupling to $\sigma_z$
for both qubits, other possibilities clearly also work. For
example, if the coupling to qubit 1 is instead proportional to
$X(\pi/2) \sigma_{x1} \equiv X(\pi/2) H_1 \sigma_{z1} H_1$ (where
$H_1$ is the Hadamard operation on qubit 1) then, subject to the
same conditions as before (local unitaries $U_i = 2^{-1/2}(1 - i
\sigma_{zi})$ applied to each qubit and a global phase of $\pi/4$),
the two-qubit gate is equivalent to CNOT rather than a
controlled-phase gate. This can easily be seen by starting with
$|\Psi_i\rangle = 2^{-1/2}(|00\rangle + |01\rangle)|\alpha\rangle$
rather than the state of Eq.~(\ref{Psiinitial}). The approach therefore has
significant flexibility in its ability to produce a universal
two-qubit gate.

\section{Specific example---superconducting charge qubits}

As a specific example we consider the case of superconducting charge
qubits \cite{shn97,mak01}. Following the original demonstration of
single charge qubit behaviour \cite{nak99}, there have been a number
of experimental demonstrations of single \cite{vio02,yu02,mar02} and
two-qubit \cite{pas03,yam03,ber03,mcd05}
charge or charge-phase qubit behaviour, culminating in the recent
demonstrations of coherent coupling to a bus mode \cite{wal05,wal04}.
So such systems certainly form a promising route for our approach.
A single charge qubit can be
thought of as a very small superconducting island---a
capacitor (of capacitance $C$)---with Josephson tunnel coupling
(of Cooper pair charges $2e$)
to a larger superconducting reservoir.
The characteristic electrostatic energy of the system is $E_c = (2e)^{2}/2$
and the Josephson tunnelling energy is $E_J$.
For qubit operation the system is biased with an
external quasi-static voltage source ($V_x$) such that the states of
the island with zero and one excess Cooper pair charge are near
degenerate, and form a good approximation to a qubit computational
basis. The size of the Josephson coupling $E_J$ can be varied externally
by creating a composite junction from two parallel junctions in a
loop threaded by a magnetic flux $\Phi_{xc}$.
\begin{figure}[!htb]
\begin{center}\includegraphics[scale=0.7]{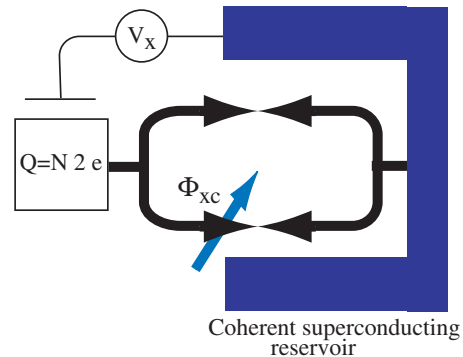}\end{center}
\caption{Schematic diagram of a charge qubit. A superconducting island with
excess charge $2Ne$ is connected to a reservoir through a composite
Josephson junction, whose effective tunnelling amplitude is controlled
by the magnetic flux $\Phi_{xc}$. An external voltage bias $V_x$ is applied to
induce an additional polarization charge.}
\label{chargequbit}
\end{figure}
Such a charge qubit is illustrated in Fig.~\ref{chargequbit}.
Putting these external sources in characteristic dimensionless terms
($n_x = CV_x/2e$, where $C$ is the effective capacitiance, and
$\phi_x = 2 \pi \Phi_{xc}/\Phi_0$) the charge qubit Hamiltonian can
be written as \cite{shn97,mak01,spil00}
\begin{equation}
H_{cq} = E_c \sigma_z (n_x - \frac{1}{2})- E_J \sigma_x \cos \phi_x\; .
\end{equation}

\begin{figure}[!htb]
\begin{center}\includegraphics[scale=0.5]{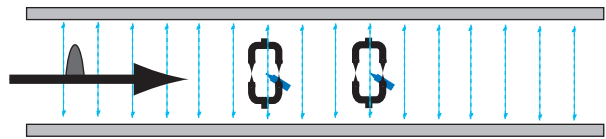}\end{center}
\caption{Schematic diagram of a two-charge-qubit system, with
each qubit coupled to the electric field of a microwave bus mode.}
\label{squid-probgate}
\end{figure}

Now, viewed as a planar structure, if such a qubit is placed in a
microwave field mode at a position where there is a non-zero
electric field (denoted as the quadrature $X(0)$) across the
capacitor and junction, there will be a microwave contribution to
$n_x$ and thus a coupling of the desired form $X(0) \sigma_z$
\cite{wal05,bla04}. Consider two such charge qubits, with no direct
coupling to each other but both positioned so they couple to the
electric field antinode of a microwave mode. This is illustrated
schematically in Fig.~\ref{squid-probgate}. The entangling gate
based on controlled displacements followed by microwave field
measurement described earlier could be applied to this
two-charge-qubit system, either in its full form, or in its
probabilistic heralded form. Given typical practical microwave
wavelengths (cm maybe down to mm), clearly many micron-scale qubits
could all be placed at the same bus field antinode, forming a useful
quantum processor or register.

\begin{figure}[!htb]
\begin{center}\includegraphics[scale=0.6]{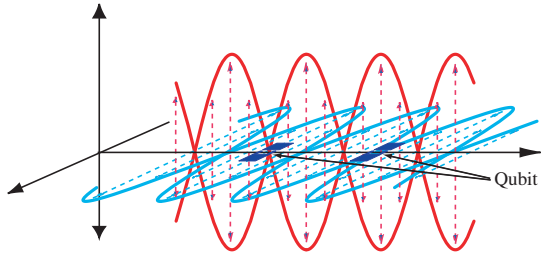}\end{center}
\caption{Schematic diagram of an array of planar charge qubits coupled to a common
microwave bus mode. Adjacent qubits are respectively coupled to the antinode
of the electric field (across the junction) and the antinode of the magnetic
field (threading the composite junction loop), enabling the forms of coupling
needed for the measurement-free gate.}
\label{twochargequbits}
\end{figure}

Alternatively, if a charge qubit is at a position where there is a
non-zero magnetic field (denoted as the quadrature $X(\pi/2)$)
normal to the plane of the structure and threading the composite
junction, there will be a microwave contribution to $\phi_x$ and
coupling of the form $X(\pi/2) \sigma_x \sin \phi_{xqs}$, which is
controllable through the quasi-static part of the field
$\phi_{xqs}$. Consider two charge qubits, one positioned to couple
to the electric field antinode of a microwave mode and the other
positioned to couple to the magnetic field antinode of the same
mode, as illustrated schematically in Fig.~\ref{twochargequbits}.
With such a system it is clearly possible to realise couplings to
different field quadratures, and thus perform a universal two-qubit
quantum gate without any post-interaction measurement of the bus
mode. As this form of gate links qubits at different field
antinodes, this provides for distributed gates between qubits
separated by cm/mm distances.

There are clearly many other possibilities that can be considered.
In terms of bus-measurement-enabled gates or interactions, different
forms of charge qubit (e.g. superconducting and semiconducting)
could be used. Gates between magnetic flux qubits
\cite{Boc97,Moo99,chi03}, all coupled to the same magnetic field
antinode of a microwave mode, could be effected through this
approach. Experimental evidence for coherent coupling between a flux
qubit and an electromagnetic oscillator has already been seen
\cite{chi04}. Gates between flux qubits and other forms of magnetic
qubit is a further possibility. In terms of measurement-free gates,
it is possible to design a new form of charge qubit (including a
$\pi$-junction) that enables geometric two-qubit gates through
interaction with a common microwave bus \cite{zhu05}. Two-qubit
interactions between charge and flux qubits (suitably positioned to
couple to the relevant microwave field quadratures) is yet another
possibility. A long term goal could therefore be a quantum computer
architecture consisting of a sizeable register (of like
qubits) suitably positioned at each microwave bus mode antinode,
functioning through "local" gates between qubits in the same
register and distributed gates between qubits in different
registers. Scaling up even further, it would be natural to consider
multiple buses, each containing a multi-register processor, all
coupled together to form a larger computer.
However, at present clearly the most immediate goal is
the experimental demonstration of local and distributed gates
between an appropriate pair of qubits, mediated through a 
single microwave bus.

\section{Two-qubit gate through controlled bus rotation and measurement}

For some solid state qubit systems, or in certain limits of
behaviour of some systems, the interaction with a bus mode takes the 
effective form of a cross-Kerr non-linearity (for instance see Ref \cite{bla04}), 
analogous to that for optical systems (see appendix 1 for details). In this case we 
have an interaction Hamiltonian of the form
\begin{equation}
\label{Hintck}
H_{int} = \hbar \chi \sigma_z a^{\dagger} a
\end{equation}
rather than that of Eq. (\ref{Hint}). When acting for a time $t$ on a
qubit-bus system, this interaction effects a rotation (in phase space)
of $\pm \theta$ on a bus coherent state, where $\theta = \chi t$ and
the sign depends on the qubit computational basis amplitude. Now it is
known already in the quantum optics context that
such interactions can be used to effect a universal two-qubit gate
between photonic qubits,
based on bus measurement \cite{nem04}. Here we give two examples
of a two-qubit parity gate, based on different forms of bus
measurement.

\begin{figure}
\begin{center}\includegraphics[scale=0.6]{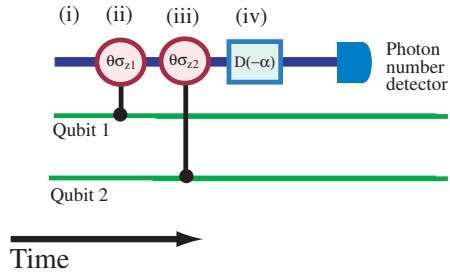}\end{center}
\caption{The circuit diagram for a two-qubit parity gate based on
controlled rotations between the qubits and the probe bus,
followed by bus number measurement.}
\label{circ-qnd-rot}
\end{figure}
\begin{figure}
\begin{center}\includegraphics[scale=0.45]{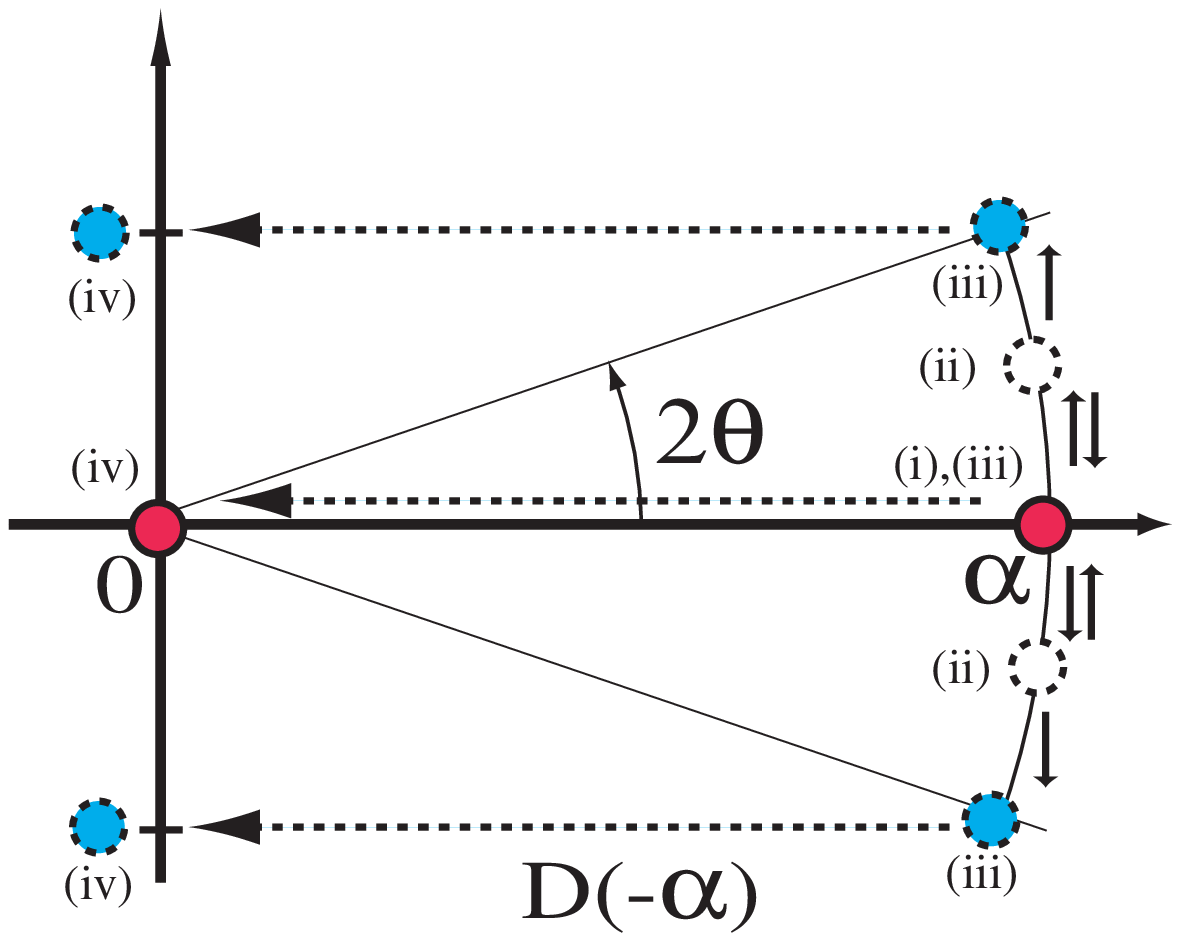}\end{center}
\caption{Schematic diagram showing the phase space evolution of the bus amplitudes
corresponding to the various qubit computational basis states. The
labels (i) to (iv) correspond to the evolution stages, as shown in the
circuit diagram in the previous figure.}
\label{qnd-rot}
\end{figure}

The circuit diagram for the first gate is shown in Fig.~\ref{circ-qnd-rot}.
Following the interactions and an {\it unconditional}
displacement operation $D(-\alpha)$, an initial
two-qubit-bus product state of Eq.~(\ref{Psiinitial}) is transformed to
\begin{eqnarray}
\label{twoqubitbusrot}
|\Psi_f\rangle = \frac{1}{2} \left( |00\rangle |\alpha (e^{2i\theta} - 1)
\rangle +  (|01\rangle + |10\rangle)|0\rangle \right. \nonumber \\
\left. + |11\rangle |\alpha (e^{-2i\theta} - 1) \rangle \right),
\end{eqnarray}
assuming equal strength coupling of both qubits to the bus mode.
This is illustrated schematically in Fig.~\ref{qnd-rot}.
A photon number measurement applied to the bus mode clearly either picks
out the vacuum, or projects onto the other two
amplitudes without distinguishing them.
Now, for $2 \theta \ll 1$ and an ideal projection
onto $|n\rangle$, the two-qubit state
is conditioned to
\begin{eqnarray}
|\psi_f\rangle = 2^{-1/2}(|01\rangle + |10\rangle)
\label{vacresultrot}
\end{eqnarray}
for $n=0$ and to
\begin{eqnarray}
|\psi_f\rangle = 2^{-1/2}(i^n |00\rangle + (-i)^n |11\rangle)
\label{nresultrot}
\end{eqnarray}
for $n>0$. These happen with equal probability of $\frac{1}{2}$
(with the most likely value of $n$ in the latter case being $n \sim
4|\alpha \theta|^2$), although there is an error probability of
$e^{-4|\alpha \theta|^2}$ (and a corresponding admixture to the state) in
the former case, due to the overlap of the coherent state with the
vacuum. This error can be made small with a suitable choice of
$\alpha$ for some given $\theta$. The phase factor in the $n>0$ case is
heralded by the
measurement outcome $n$, and so can be allowed or corrected for.
Clearly using a photon number measurement this gate is near-deterministic. Alternatively
one can also use the iterative procedure with the intensity measurement/bucket
detector described in Section (III) to enable near-deterministic entanglement generation.
If one is instead prepared to accept a probabilistic gate, heralded by the measurement
outcome, then a measurement of the $X(\pi/2)$ quadrature of the bus can be used.
Half the time it will project to Eq.~(\ref{vacresultrot}), heralded by a result
close to zero. The other half of the time no entanglement will be produced. Such
a procedure may be a good approach for initial experimental demonstrations of the
principle. It would not require the final unconditional displacement.

\begin{figure}[!ttb]
\begin{center}\includegraphics[scale=0.6]{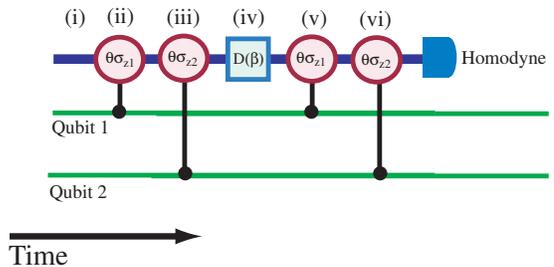}\end{center}
%\begin{center}\includegraphics[scale=0.5]{displacement}\end{center}
\caption{The circuit diagram for a two-qubit parity gate based on
controlled rotations between the qubits and the probe bus,
followed by bus $X(0)$ quadrature measurement. The amplitude of the unconditional
displacement is given by $\beta=-2 \alpha \cos 2 \theta$.}
\label{circ-par-rot}
\end{figure}

\begin{figure}[!ttb]
\begin{center}\includegraphics[scale=0.4]{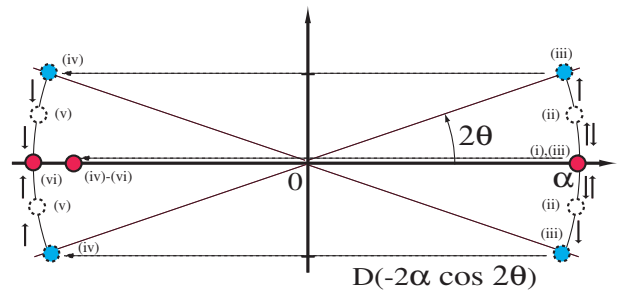}\end{center}
\caption{Schematic diagram showing the phase space evolution of the bus amplitudes
corresponding to the various qubit computational basis states. The
labels (i) to (vi) correspond to the evolution stages, as shown in the
circuit diagram in the previous figure.}
\label{par-rot}
\end{figure}

A near-deterministic gate based on a single final quadrature measurement can be
achieved, although it requires a more involved circuit.
This is shown in Fig.~\ref{circ-par-rot}.
Following qubit-bus interactions analogous to the previous gate
and an {\it unconditional}
displacement operation $D(- 2 \alpha \cos 2 \theta)$,
further qubit-bus interactions occur. An initial
two-qubit-bus product state of Eq.~(\ref{Psiinitial})
 is transformed to
\begin{eqnarray}
\label{twoqubitbusrotXmeas}
|\Psi_f\rangle &=&  \frac{1}{2} \left[ (|00\rangle + |11\rangle)|- \alpha
\rangle \right. \nonumber \\
 &\;&\;\;\;\;+   \left. (|01\rangle + |10\rangle)|\alpha (1 - 2 \cos 2 \theta) \rangle \right] ,
\end{eqnarray}
again assuming equal strength coupling of both qubits to the bus mode.
This is illustrated schematically in Fig.~\ref{par-rot}.
Clearly with this scheme a homodyne measurement of the $X(0)$ quadrature
of the bus mode will project onto the odd or even parity entangled
two-qubit states that sit in Eq.~(\ref{twoqubitbusrotXmeas}), with the outcome
heralded by the quadrature result. This parity gate isn't perfect, as
the final states of the CV bus mode corresponding to the
different entangled states of qubits
are not exactly orthogonal. (See Fig.~\ref{error} for an illustration.)
However, as with the qubit measurement scenario,
for the example of taking the midpoint between the
probability peaks as the discrimination point and using $X(0)$
quadrature measurement of the bus, the error probability (the sum of
the areas that sit the wrong side of the discrimination point) is approximately
$E = \frac{1}{2}$erfc$(2^{1/2} |\alpha| \theta^2)$. This can be made very small
for a suitable choice of $\alpha \theta^2$.  As in the qubit measurement case,
somewhat imperfect homodyne measurement can be tolerated provided that
$\alpha \theta^2$ is large enough to dominate the homodyne error.


\section{Two-qubit gate through controlled bus rotations alone}

\begin{figure}
\begin{center}\includegraphics[scale=0.7]{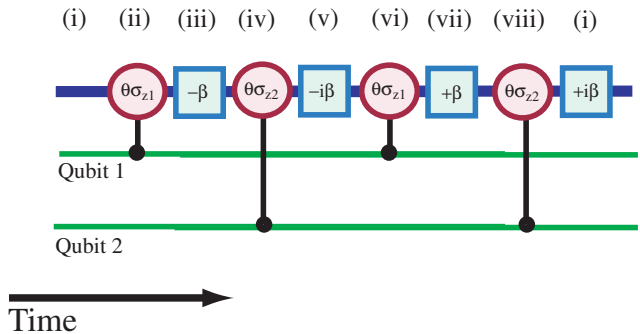}\end{center}
\caption{The circuit diagram of a two-qubit controlled-phase gate,
based on controlled rotations between the qubits and the probe bus
and non-controlled displacements of the bus.}
\label{circ-rot}
\end{figure}

With cross-Kerr interactions of the form of Eq.~(\ref{Hintck}) it is also
possible to achieve a
universal two-qubit gate without any bus measurement.
The relevant gate sequence is illustrated in Fig.~\ref{circ-rot}.

\begin{figure}
\begin{center}\includegraphics[scale=0.4]{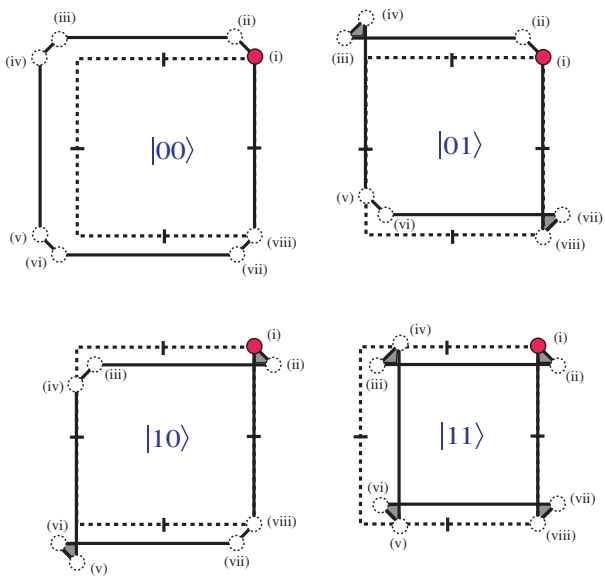}\end{center}
\caption{Schematic diagram showing the phase space evolution of the bus amplitudes
corresponding to the various qubit computational basis states. The
labels (i) to (viii) correspond to the evolution stages, as shown in the
circuit diagram in the previous figure. The small areas shown in
grey are traversed in a
clockwise sense, giving a negative contribution to the phase acquired.}
\label{rot}
\end{figure}

In this case the unconditional displacements are all of equal
magnitude $\beta$ (but varying directions as shown in Fig.~\ref{circ-rot}) and
the controlled rotations are generated through Hamiltonians of the
form (\ref{Hintck}) with an interaction time $t$ and $\theta=\chi
t$. With an initial state of (\ref{Psiinitial}) and $\alpha = \beta
(1+i)/2$, the gate sequence of Fig.~\ref{circ-rot} achieves a gate which is
locally equivalent to a controlled-phase gate for the condition
$|\beta \theta|^2 = \pi/4$.
The behaviour of the bus amplitudes in phase space is illustrated in
Fig.~\ref{rot}. There are a number of points that should be noted about
this gate.

\begin{enumerate}
\item  There are simpler circuits available for the two-qubit gate
through controlled bus rotations alone.
These use the same number of controlled rotations but fewer
probe bus displacement operations. For example, in the circuit given
in Fig.~\ref{circ-rot}, the final displacement is not actually necessary to
implement the gate. In this case the final displacement just returns the
probe beam to its initial
starting position, which is neat but unnecessary.
\item There are also variations of these gates based on two displacement operations.
Unfortunately,
while performing a CNOT or CPhase, these do not have the same
scaling in  terms of $\theta$ and $\beta$
for the resultant phase shift.
\item In the rotation-based case of Fig.~\ref{rot}, unlike that of the
displacement-based gate of Fig.~\ref{displacements}, there is an error.
The bus mode doesn't
disentangle from the qubits exactly, because the (small) rotations
employed are arcs of circles, rather than straight lines. The error is of
order $|\beta \theta^2|$, which can clearly be made small (of order $1/\beta$) even
for a maximally entangling universal gate by working in the small
$\theta$ large $\beta$ limit. The gate is discussed in more detail in
the Appendix 2.
\end{enumerate}

\section{Specific examples}

It is already known theoretically \cite{bla04} that a superconducting charge
qubit (as illustrated in Fig.~\ref{chargequbit}) coupled to a microwave
field mode in the dispersive limit has an interaction Hamiltonian of the
cross-Kerr form of Eq.~(\ref{Hintck}). Furthermore, experiments in this limit
\cite{wal04,wal05} have clearly demonstrated this coupling. Such
systems are clearly excellent candidates for the controlled-rotation-based
gates that we propose. Two charge qubits coupled to a microwave field mode
(as illustrated in Fig.~\ref{squid-probgate})
in the dispersive limit form a candidate system for realising the
measurement-based gates. The simplest initial demonstration would probably
be to employ the scheme given in Fig.~\ref{circ-qnd-rot} in a probabilistic
approach, in which case the final non-controlled displacement is unnecessary.
Measurement of the $X(\pi/2)$ quadrature of the microwave field would
entangle the charge qubits with probability 1/2, heralded by a measurement outcome
close to zero. If such a gate could be achieved, this would point the way
towards the near deterministic gates of Figs.~\ref{circ-qnd-rot} and
\ref{circ-par-rot}, based respectively on photon number or $X(0)$
quadrature measurement of the microwave field. With sufficient control over
the couplings and the microwave field, the geometric gate of Fig.~\ref{circ-rot}
between two dispersive charge qubits is a further possibility.

\begin{figure}[!htb]
\begin{center}\includegraphics[scale=0.4]{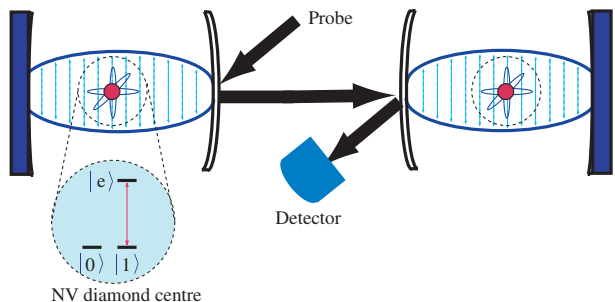}\end{center}
\caption{Schematic diagram of a two NV-diamond centre qubits in separate
cavities coupled via a cross-kerr interaction to an optical field. The final
measurement enables the construction of a parity gate. .}
\label{nv}
\end{figure}

We can also consider NV-diamond centres \cite{martin99,jelezko04} within individual cavities as
excellent matter system candidates for the qubus protocol and especially the parity gates
(as depicted in Fig ~\ref{nv}). Within the level structure of an NV-diamond centre are
two long-lived states
$|{\uparrow}\rangle$ and $|{\downarrow}\rangle$ and an excited state
$|e\rangle$, in an $L$-configuration with the $|{\downarrow}\rangle \leftrightarrow |e\rangle$
transition coupled to the cavity mode. The $|{\uparrow}\rangle$ state can represent
the logical $|0\rangle$ basis state and the $|{\downarrow}\rangle$ state the logical
$|1\rangle$ basis state. The NV diamond level configuration is such that only the
$|1\rangle$ state can be excited to the state $|e\rangle$ via an optical
pulse while the $|0\rangle$ to $|e\rangle$ transition is assumed forbidden or
extremely weak. When the optical pulse interacts with the $|1\rangle$ state, it picks up
a small phase shift due to the $|1\rangle \leftrightarrow |e\rangle$ transition
while no phase shift occurs for the $|0\rangle$ state. This essential difference is
all that we require to create a conditional phase shift  and thus implement the two qubit gates
through controlled rotation.

\section{Controlled displacements from controlled rotations}

The previous sections have demonstrated the power of qubit-controlled displacements
and rotations, applied to a communication bus, for fundamental
two-qubit gate operations and quantum information processing. The
controlled displacement gates seem potentially easier to implement,
as they do not require unconditional displacements as well and operate for
small to large values of $\chi t$. However the controlled displacement schemes
generally have a strong dipole coupling requirement, needed to ensure that $\omega t$
can be assumed constant through out the gate. For SQUID and other microwave based
schemes this condition can be satisfied easily, but it poses a significant problem
in the optical regime, suggesting that optical schemes are restricted to
controlled rotation based gates. However, this is not the case,
as it is straightforward to transform a controlled rotation interaction
to a controlled displacement. Consider the Hamiltonian in the interaction
picture given by Eq.~(\ref{Hintck}) with the bus mode displaced by an amount
$\alpha$. In this case
\begin{equation}
\hbar \chi \sigma_z a^{\dagger} a \rightarrow \hbar \chi \sigma_z \left[|\alpha|^2
+ \alpha^\ast a^{\dagger}+ \alpha a+a^{\dagger} a\right] \; .
\end{equation}
If we let $\alpha=|\alpha|e^{-i \theta}$ then the above equation can be written in
the form $H_{int}=\hbar \chi \sigma_z \left[|\alpha|^2+ |\alpha| X(\theta)+a^{\dagger} a\right]$.
We clearly see a controlled displacement term, plus two other pieces dependent
on $a^{\dagger} a$ and $|\alpha|^2$.
These terms can be eliminated with a simple trick as follows.
First run the interaction given by
$U(\alpha,\sigma_z,t)=\exp \left[ i H_{int} t / \hbar\right]$ for
a time $t/4$, then bit-flip the qubit and change the sign of the displacement
from $+\alpha \rightarrow -\alpha$ and run for a further time $t/2$. Finally repeat
the orginal $U(\alpha,\sigma_z,t)$ for a time $t/4$. This results in a net interaction
$U(\alpha,\sigma_z,t/4) U(-\alpha,-\sigma_z,t/2) U(\alpha,\sigma_z,t/4) \sim
\exp \left[ i  |\alpha| \chi t \sigma_z X(\theta')+O[t^3,a,a^\dagger]\right]$
which is the desired displacement. There is now
an effective coupling constant  $|\alpha| \chi$, where $\alpha$ can in principle
be large.  There is a small $O(t^3)$ correction in the above evolution but this is tiny
for reasonable interaction times. The key issue becomes how to achieve
the displacement on the probe field. There are a number of well known
solutions to this but the easiest is to drive the probe field with a classical pump
where the displacement is required. This is experimentally achievable.

It is worthwhile considering a specific situation in a litte more detail. Consider a
Lambda-type atomic configuration with the two lowest energy levels representing the logical
qubit states. We know that if we detune by $\Delta$ a field $a$ that connects the logical
$|1\rangle$ to the excited state $|e\rangle$ we get an effective coupling
$H_I= \hbar \frac{g^2}{2\Delta} a^\dagger a |1\rangle \langle 1|$, where $g$ is the coupling
coefficient between the atomic system and probe mode. A second field $b$ detuned above the
$|e\rangle$ by the same amount gives
$H_I=- \hbar\frac{g^2}{2\Delta} b^\dagger b |1\rangle \langle 1|$.
Now if we choose the field $a$ to have a large classical component,
then $a\rightarrow a + |\alpha| e^{-i \theta}$
and the field $b$ to be pure classical $b \rightarrow|\alpha| e^{-i \phi}$
then the two effective Hamiltonian yields a net Hamiltonian
\begin{equation}
H_{eff}= \hbar \frac{g^2}{2\Delta}\left[ |\alpha| X(\theta)+a^\dagger a \right]|1\rangle \langle 1|
\end{equation}
where the quadratic Stark shift of the level $|1\rangle$ is cancelled and there is no special
phase relation required between the upper and lower detunings. Lastly for $|\alpha|$ large the
component $a^\dagger a|1\rangle \langle 1|$ can be neglected or eliminated as discussed above.
There are many variations of this approach---the one to be used in practice depends upon the
actual experimental system.

\section{A further example based on rotations and displacements}

\begin{figure}[!htb]
\begin{center}\includegraphics[scale=0.7]{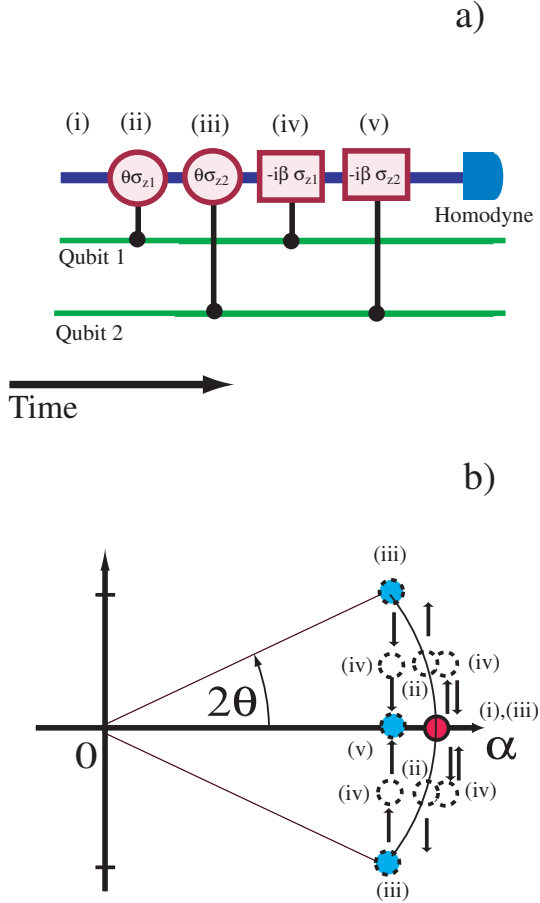}\end{center}
\caption{a) The circuit diagram of a two-qubit parity gate,
based on controlled rotations and controlled displacements
between the qubits and the probe bus.
b) Schematic diagram showing the phase space evolution of the bus amplitudes
corresponding to the various qubit computational basis states. The
labels (i) to (v) correspond to the evolution stages, as shown in the
circuit diagram in a).}
\label{dis-rot}
\end{figure}

Finally, we present a near-deterministic gate based on a final quadrature measurement and
with qubits that are able to both rotate and displace the bus mode.
This may be rather more difficult to achieve for the current most popular
matter qubits, in comparison to the gates already presented, but we include
it to cover the full spectrum of possibilities.
The circuit diagram is shown in Fig.~\ref{dis-rot}.
Following this conditional gate sequence, an initial
two-qubit-bus product state of Eq.~(\ref{Psiinitial})
 is transformed to
\begin{eqnarray}
\label{twoqubitdisrotXmeas}
|\Psi_f\rangle = \frac{1}{2} \left( (|00\rangle + |11\rangle)|\alpha
(1 - \cos 2 \theta) \rangle \right. \nonumber \\
\left. +  (|01\rangle + |10\rangle)|\alpha \rangle \right),
\end{eqnarray}
assuming equal strength coupling of both qubits to the bus mode
and $\beta = \frac{\alpha}{2} \sin 2 \theta$.
The phase space evolution of the bus mode amplitudes is
illustrated schematically in Fig.~\ref{par-rot}.
The result is two amplitudes lying on the real axis, so
a homodyne measurement of the $X(0)$ quadrature
of the bus mode will project onto the odd or even parity entangled
two-qubit states that sit in Eq.~(\ref{twoqubitdisrotXmeas}), with the outcome
heralded by the quadrature result. As with some previous examples,
this parity gate isn't perfect, as the final states of the CV bus mode
corresponding to the different entangled states of qubits
are not exactly orthogonal, as illustrated in Fig.~\ref{error}.
Once again, taking the midpoint between the
probability peaks as the discrimination point, the error probability (the sum of
the areas that sit the wrong side of the discrimination point) is approximately
$E = \frac{1}{2}$erfc$(2^{-1/2} |\alpha| \theta^2)$. This can be made very small
for a suitable choice of $\alpha \theta^2$ and
somewhat imperfect homodyne measurement can be tolerated provided that
$\alpha \theta^2$ is large enough to dominate the homodyne error.

\section{Discussion}

We have presented a new approach to quantum computing---a ``qubus
computer''---which brings together discrete qubits with quantum
continuous variables in a single scheme. Through interaction with a
common bus mode, it is possible to realise a universal two-qubit
gate. We considered three different schemes including:
\begin{itemize}
\item Measurement-based probabilistic but heralded parity gates,
\item Measurement-based near deterministic parity gates and
\item Measurement-free deterministic CPhase gates,
\end{itemize}
with two different interactions (the controlled-displacement and the
controlled-rotation) between the discrete qubits and the bus mode.
For the latter scheme, no post-interaction measurement is required
on the bus mode --- it effectively plays the role of a catalyst in
enabling the gate. All of these approaches are particularly well suited
for solid state qubits, which generally have a natural dipole
coupling to a common electromagnetic  field mode, such as superconducting
qubits coupled to a microwave field or an NV diamond centre coupled to an
optical cavity mode. However the results are also directly applicable to all
optical gates.

Lastly, our approach does not generally force a choice of computation scheme and
processor architecture; rather it provides building blocks which can
be put together to suit the task at hand. For instance, the near deterministic
gates can be used for the standard gate-based quantum computation as well
computation by measurement (the one-way quantum
computer \cite{Briegel01,Nielsen04,Browne04} for instance) or the simulation
of Hamiltonians. Our approach requires only a practical set of resources, and it
uses these very efficiently. Thus it promises to be extremely useful for
the first quantum technologies, based on scarce resources. Furthermore,
in the longer term this approach provides both options and scalability
for efficient many-qubit quantum computation.

\noindent {\em Acknowledgments}: We thank R. Van Meter, S. D. Barrett, R. G. Beausoleil,
P. Kok, T. Ladd and P. L. Knight for valuable discussions. This work was supported in  part
by the Japanese JSPS, MPHPT, and Asahi-Glass research grants, the UK research council EPSRC,
the Australian Research Council Centre of Excellence in Quantum
Computer Technology and the European Project RAMBOQ. SLB currently holds a
Royal Society Wolfson Research Merit Award.

\section{Appendix 1}
Consider the three level Raman coupling scheme in a Lambda configuration. A strong coherent 
field is detuned from the dipole allowed transition $|0\rangle\leftrightarrow|e\rangle$, 
while a weaker cavity field (annihilation operator $a$) is detuned by an equal amount 
from the dipole transition $|1\rangle\leftrightarrow|e\rangle$. The effective two level 
Hamiltonian describing this Raman process is\cite{Leibfried}
\begin{equation}
H=\hbar\frac{\omega_0}{2}\sigma_z+\hbar\omega_c a^\dagger a+
\frac{\hbar\Omega}{2}(a\sigma_++a^\dagger \sigma_-)
\end{equation}
where $\sigma_z=|0\rangle\langle 0|-|1\rangle\langle 1|,\ \ \  \sigma_+=|1\rangle\langle 0|$ 
and $\hbar\omega_0$ is the energy difference between the qubit states while $\omega_c$ is 
the frequency of the quantised cavity mode that acts as the quantum bus.

We now transform to an interaction picture by the unitary operator
\begin{equation}
U_o(t)=\exp[-i\frac{\omega_0}{2} \sigma_zt-i\omega_c a^\dagger a t] 
\end{equation}
so that the state in the interaction picture satisfies
\begin{equation}
\frac{d|\psi(t)\rangle_I}{dt}=-\frac{i}{\hbar}H_I(t) |\psi(t)\rangle_I
\end{equation}
where 
\begin{equation}
H_I(t)=\frac{\hbar\Omega}{2}(a\sigma_+e^{-i\Delta t}+a^\dagger \sigma_+ e^{i\Delta t})
\end{equation}
and $\Delta=\omega_c-\omega_0$. 
The solution is
\begin{eqnarray}
|\psi(t)\rangle_I&=&\left[1-\frac{i}{\hbar}\int_0^t dt_1 H_I(t_1)\right.  \\
&\;&\left. +\frac{1}{2}(\frac{-i}{\hbar})^2\int_0^t dt_2\int_0^{t_2} dt_1 H_I(t_2)H_I(t_1)+\ldots\right] \nonumber
\end{eqnarray}

For sufficiently large detuning,  the relevant time scale of the dynamics is 
such that $\Delta t>>1$. We can make the following approximations
\begin{eqnarray*}
\int_0^t dt_1H_I(t_1)& \approx &  0\\
\int_0^tdt_2\int_0^{t_2} dt_1 H_I(t_2)H_I(t_1) & \approx & -i\frac{\hbar^2\Omega^2}{4\Delta}(aa^\dagger \sigma_+\sigma_--a^\dagger a\sigma_-\sigma_+)
\end{eqnarray*}
Continuing in this way we can find that all terms of odd order in $\Omega$ 
can be neglected, while all terms of order $\Omega^{2n}$ are proportional to $t^{2n}$, so that the dynamics can be approximated by
\begin{equation}
|\psi(t)\rangle_I=e^{-i H_{eff} t/\hbar}
\end{equation}
where, by using the commutation relations, the effective interaction 
Hamiltonian can be written as 
\begin{equation}
H_{eff}=\hbar\chi a^\dagger a \sigma_z
\end{equation}
There is also a small renormalisation of the atomic and cavity frequencies that we have ignored. 

\section{Appendix 2}

It is worthwhile examining the two-qubit gate through controlled bus
rotations alone in a little more detail, as there are a number of
subtle issues in its operation. We will assume that the probe bus starts initially
in the state $|\alpha e^{i \frac{\pi}{4}} \rangle$ with $\alpha$ real. The
displacement is given $\beta=\sqrt{2} \alpha$. It is straighforward
but tedious then to show that the four basis states (including the
probe bus) evolve as
\begin{eqnarray}
|g \rangle |g \rangle |\alpha e^{i \frac{\pi}{4}} \rangle &\rightarrow& e^{i \phi_{gg}} |g \rangle |g \rangle |\alpha_+ \rangle \\
|g \rangle |e \rangle |\alpha e^{i \frac{\pi}{4}} \rangle &\rightarrow& e^{i \phi_{ge}} |g \rangle |e \rangle |\alpha e^{i \frac{\pi}{4}} \rangle \\
|e \rangle |g \rangle |\alpha e^{i \frac{\pi}{4}} \rangle &\rightarrow& e^{i \phi_{eg}} |e \rangle |g \rangle |\alpha e^{i \frac{\pi}{4}}\rangle \\
|e \rangle |e \rangle |\alpha e^{i \frac{\pi}{4}} \rangle
&\rightarrow& e^{i \phi_{ee}} |e \rangle |e \rangle |\alpha_-\rangle
\end{eqnarray}
where the phase shift $\phi_{gg},\phi_{ge},\phi_{eg},\phi_{ee}$ are given by
\begin{eqnarray}
\phi_{gg}&=&  \alpha^2 \left[7 \cos \theta-\cos 2 \theta- 3 \cos 3\theta + \cos 4 \theta\right] \nonumber \\
&\;& \;\;\;\;+  \alpha^2 \left[ \sin \theta + 5\sin 2 \theta - \sin 3 \theta+\sin 4 \theta \right]\\
\phi_{ge}&=&  4 \alpha^2 \cos \theta \\
\phi_{eg}&=&  4 \alpha^2 \cos \theta \\
\phi_{ee}&=&  \alpha^2 \left[7 \cos \theta-\cos 2 \theta- 3 \cos 3\theta + \cos 4 \theta\right] \nonumber \\
&\;& \;\;\;\;- \alpha^2 \left[ \sin \theta + 5\sin 2 \theta - \sin 3 \theta+\sin 4 \theta \right]
\end{eqnarray}
and the amplitude $\alpha_\pm$ of probe bus states $|\alpha_\pm\rangle$ are
\begin{eqnarray}
\alpha_\pm =\alpha \left(e^{\pm 4 i \theta+i \frac{\pi}{4}}+\sqrt{2}\left[1-e^{\pm 2 i \theta}\right]\left[i+e^{\pm i \theta}\right] \right) \end{eqnarray}
We immediately notice that the probe bus for the $|g \rangle |g
\rangle$ and $|e \rangle |e \rangle$ basis state has not returned
exactly to the initial starting point $|\alpha e^{i \frac{\pi}{4}}\rangle$.
Instead they have returned to the states $|\alpha_+ \rangle$ and
$|\alpha_- \rangle$ respectively for the basis qubit states $|g
\rangle |g \rangle$ and $|e \rangle |e \rangle$. One can think of these probe bus
states as being slightly displaced from $|\alpha e^{i \frac{\pi}{4}}\rangle$.
Ignoring the probe bus will then introduce decoherence in the matter qubits (in fact a dephasing effect).
Tracing out the probe bus we get
\begin{eqnarray}
|g \rangle |g \rangle &\rightarrow& {\hat {\cal P}}[e^{- \gamma_{gg}}] e^{i \phi_{gg}+i  \psi_{gg}} |g \rangle |g \rangle \\
|g \rangle |e \rangle &\rightarrow& e^{i \phi_{ge}} |g \rangle |e \rangle \\
|e \rangle |g \rangle &\rightarrow& e^{i \phi_{eg}} |e \rangle |g \rangle \\
|e \rangle |e \rangle &\rightarrow&  {\hat {\cal P}}[e^{-
\gamma_{ee}}] e^{i \phi_{ee}+i \psi_{ee} } |e \rangle |e \rangle
\end{eqnarray}
where
\begin{eqnarray}
\psi_{gg}&=& -4 \alpha^2 \sin^2 \frac{\theta}{2} \left(1+\cos \theta+\sin \theta \right) \nonumber \\
&\;&\;\;\;\;\; \times \left(\cos 2\theta+\sin 2\theta \right) \\
\psi_{ee}&=& -4 \alpha^2 \sin^2 \frac{\theta}{2} \left(1+\cos \theta-\sin \theta \right) \nonumber \\
&\;&\;\;\;\;\; \times \left(\cos 2\theta-\sin 2\theta \right)
\end{eqnarray}
and $ {\hat {\cal P}}$ is the dephasing projector with
\begin{eqnarray}
\gamma_{gg}&=& 16 \alpha^2 \sin^4 \frac{\theta}{2}
\left(1+\cos \theta-\sin \theta \right)^2\\
\gamma_{ee}&=& 16 \alpha^2 \sin^4 \frac{\theta}{2}
\left(1+\cos \theta+\sin \theta \right)^2
\end{eqnarray}
It is now very clear that tracing out the probe bus has resulted in
an extra phase shift on the $|g \rangle |g \rangle$ and $|e \rangle
|e \rangle$ basis states as well a dephasing term. As long as
$\theta \ll 1$, $\gamma_{gg} \sim 4 \alpha^2 \theta^4 \ll 1$ and so
has a negligible effects. Removing a global phase factor and
performing several local rotations, our basis qubits evolve as
\begin{eqnarray}
|g \rangle |g \rangle &\rightarrow& e^{i \phi_{d}/2 }|g \rangle |g \rangle \\
|g \rangle |e \rangle &\rightarrow& |g \rangle |e \rangle  \\
|e \rangle |g \rangle &\rightarrow& |e \rangle |g \rangle \\
|e \rangle |e \rangle &\rightarrow& e^{i \phi_{d}/2 } |e \rangle
|e \rangle
\end{eqnarray}
Setting
$\phi_{d}=\phi_{gg}-\phi_{ge}-\phi_{eg}+\phi_{ee}+\psi_{gg}+\psi_{ee}$.
It is straightforward from the above expression to show
\begin{eqnarray}
\phi_{d}= 8 \alpha^2 \sin^2 \theta  \left(2 \cos \theta-\cos
2\theta \right) \sim 8 \alpha^2 \sin^2 \theta
\end{eqnarray}
Without taking into account the effect that the probe bus was
slightly displaced for $|g \rangle |g \rangle$ and $|e \rangle |e
\rangle$ from $|\alpha \rangle$ our resultant phase shift
$\phi_{d} \sim 8 \alpha^2 \sin^2 \theta$. It is also important to mention
that the single qubit phase operations scale as $\phi_{s} \sim \alpha^2 \theta$.


\begin{thebibliography}{99}
\bibitem{inter-pic} Of course the full Hamiltonian will also contain a free evolution term of the
form $H_{0}= \hbar \Omega \sigma_z+\hbar \omega a^\dagger a$
which must be taken into account. In the
interaction picture this will give us the effective Hamiltonian
$H_{int} = \hbar \chi \sigma_z \left[ e^{i \omega t} a^\dag  + e^{-i \omega t} a\right]$.
In the situation where one has a large dipole moment ($\chi \gg \omega$), we can work
in the regime $\omega t$  constant and so $H_{int}$ is effectively the time independent
Hamiltonian $H_{int} = \hbar \chi \sigma_z X(\theta)$ where $\theta=\omega t$. By varying
$t$ in a controlled fashion we have a mechanism to change $X(\theta)$ from
the position quadrature $X(0)$  to the momentum quadrature $X(\pi/2)$.
\bibitem{projection} Such a projection will require a photon number resolving detector.
\bibitem{bucketdet} The so-called bucket detector measures whether or not the state is in
the vacuum. It can be modelled by the projectors $|0\rangle \langle 0|$ and
$I-|0\rangle \langle 0|$. For a non-zero photon number it gives no information about what
the photon number actually is. However, in many realistic situations the detector
does have some photon-number resolving characteristics and it can be modelled by the
POVM for detecting $n$ photons as $\eta^n \sum_{k=n}^\infty \left( {k \atop n }\right)
(1-\eta)^{k-n} |k\rangle \langle k|$ where $\eta$ is the efficiency of the detector.
Clearly obtaining some information about the photon number helps in determining whether
$n$ is even or odd and hence whether the system is in the state $|00\rangle + |11\rangle$ or
$|00\rangle - |11\rangle$. This changes the constant $A$ away from 1/2. This additional
coherence in the final state could be utilized (with single qubit operations)
to speed up the removal of mixture in an iterative entangling process,
compared to the simple approach
presented here.
\end{thebibliography}
\end{document}